\def\n{\noindent\/}
\def\be{\begin{equation}}
\def\ee{\nonumber\end{equation}}

\def\ggg{\!\!\gg\/}
\def\lll{\ll\!\!\/}
\documentclass[epj]{svjour} 

\usepackage{graphics}

\begin{document}
\title{Recursive approach to study transport properties of atomic wire}
\author{S. Datta \inst{1} \and T. Saha-Dasgupta \inst{1,2} \and A. Mookerjee \inst{1,2,3}
}
\institute{Department of Materials Science, S. N. Bose National Centre for Basic Sciences,
 Kolkata 700 098, India \and Advanced Materials Research Unit, S. N. Bose National Centre 
for Basic Sciences, Kolkata 700 098, India \and Unit for Nanoscience and Technology,
 S. N. Bose National Centre for Basic Sciences, Kolkata 700 098, India.}

\date{Received: date / Revised version: date}

\abstract{
In this study, we propose a recursive approach to study the transport properties of atomic wires.
 It is based upon a real-space block-recursion technique with
 Landauer's formula being used to express the conductance as a scattering problem. To illustrate
 the method, we have applied it on a model system described by a single band tight-binding 
Hamiltonian. Results of our calculation therefore may be compared with the reported results 
on Na-atom wire. Upon tuning the tight-binding parameters, we can distinctly identify the controlling
 parameters responsible to decide the width as well as the phase of odd-even oscillations in the
 conductance.
\PACS{
      {73.63.-b}{Electronic transport in nanoscale materials and structures}   \and
      {71.15.-m}{Methods of electronic structure calculation} \and
      {85.35.-p}{Nanoelectronic devices}
     } 
} 

\maketitle

\section{\label{sec:intro}Introduction}
A mono-atomic quantum wire has a cross-section of one atom and is several 
atoms long. Such a system can be formed by pulling atomic contacts
using a scanning tunneling microscope (STM) or a mechanically controllable break
junction (MCBJ). The experimental evidences of formation of such atomic wires
 have been reported  by Yanson {\it et al} \cite{1} and  Ohnishi {\it et al} \cite{add1}.
  However, issues involving electronic structure, transport properties and
 the influence of contacts with macroscopic conductors are still not fully settled. 
The most striking feature of such a mono-atomic wire is the non-monotonic behavior of
 its conductance as a function of the number of atoms along its length. Oscillatory behavior 
of conductance as a function of the length of the wire has been experimentally observed 
in Al, Pt and Ir wires by Smit {\it et al} \cite{2}, in Au and Ag wires by Thijssen 
 {\it et al} \cite{add2}, in Na wires  by Krans {\it et al} \cite{ex1}  and  in  wires of atoms of
 higher valency by Yanson {\it et al} \cite{1}. This oscillatory  behavior has been attributed
 to interference effect resulting from the changes in the connection between
the wire  and the edge of the electrode when new atoms are  pulled into
the wire.

Theoretically, the conductance of atomic wires can be calculated using the Landauer formalism in which one
 relates the conductance $G$ in the linear response regime to the transmittance at the Fermi energy, $T(E_F)$, as
~: $G \ =\ (e^2/\pi\hbar)T(E_F)$. There are several numerical methods to calculate transmittance. They
 may be broadly classified into wavefunction and Green function methods. In wavefunction 
technique, one solves for scattering wavefunction of the system using methods like transfer matrix 
method \cite{huc1,huc2,trfm1,trfm2,trfm3}, finite difference method \cite{fdm} or by solving Lippman-Schwinger equation \cite{lse1,lse2}. All these methods use various level of approximations to describe the electronic structure of the system, starting from semi-empirical models like extended H\"uckel \cite{huc1,huc2} to fully atomistic descriptions based on the density functional theory (DFT) of Kohn and Sham. In most of the wave
 function methods, however, the electronic structure of the scattering
 region is resolved in detail, while the leads are modeled by a free electron gas \cite{trfm1,trfm2,lse1,lse2}. 
Alternatively, conductance can be calculated by Green's function method which does not require 
the explicit calculation of the scattering wave function. The main effort in this approach is to calculate 
the Green function of the central wire in the presence of the coupling to the leads. The effect
 of coupling to the leads is taken into account through self-energy terms. Various methods based on this approach, mainly differ in the choice of basis set used to represent the Hamiltonian and 
self-energy matrix, e.g Gaussians \cite{gs}, numerical atomic orbitals \cite{lcao}, wavelets \cite{wavelet}
 and plane wave basis \cite{wannier}. Although implementations within the Green
 function as well as the wavefunction approaches have been carried forward using various 
different techniques, it can be shown that the approaches are completely equivalent for non-interacting
 electrons\cite{10}. In this paper we propose a combination of the scalar
 and vector recursion techniques \cite{26a,27,26} as a viable and efficient means to study 
the transport properties within the general scheme of wavefunction approach of a 
lead-wire-lead system.

As a first step, starting from the junction point of the leads and the wire, we have used the
recursion method of Haydock {\it et al}  \cite{26a,27} to map the quasi-1D semi-infinite leads with finite 
extensions along lateral directions onto equivalent chains. In the process, the information of the shape of
 the leads is encoded in the recursion coefficients. The recursion coefficients converge and the asymptotic
 part of the chains (or terminators)
 resembles periodic one-dimensional leads. This naturally divides the whole system unambiguously into an
effective scattering region and two attached ballistic, periodic chain leads. The initial variation of
the recursion coefficients contributes to the scattering region.

In the second step, the scattering matrix has been calculated by solving the Schr\"odinger equation using the vector or block recursion method of Godin and Haydock  \cite{26} and applying the wave-function matching conditions at the lead-wire interfaces.

In the final step, we have applied Landauer's formula to express the conductance of
the quantum wire.

The advantage of the proposed method is that it is a real space based method. Moreover, to solve the scattering problem, one does not need to calculate the wave function explicitly. By putting the boundary conditions at the lead-wire junctions and at the end of vector chain, one can directly calculate the scattering matrix.
The basic inputs in the above described procedure are the tight-binding (TB) Hamiltonians for the wire and the leads. In order to illustrate the method we have applied it to a model system described by a single-band TB Hamiltonian. We demonstrate the validity of this approach by comparing the results obtained out of a simple model system with the reported results on monoatomic Na-wire calculated from first-principles. Moreover, by tuning the TB parameters of the wire and the leads we have  provided a detailed understanding of the various features of the problem beyond what has been reported earlier \cite{3,5,6,7,8,ex6,ex2,ex5,ex7}. The proposed technique being based on recursion, relies on the sparseness of the starting Hamiltonian. The coupling of this method to fully atomistic 
DFT description of the electronic structure, however, can be easily achieved in terms of Wannier function based Hamiltonian constructed out of DFT calculations.

\section{\label{sec:methodology} Method}

\subsection{\label{sec:step1}Converting quasi-1D lead to effective 1D chain and detection of effective scattering
region}

Each lead with finite lateral dimension of $N_x \times N_y$ is described by a Hamiltonian in a tight-binding basis $\{\vert i\mu\rangle\}$ where
$i$ labels a site and $\mu$ a particular channel, e.g. in a linear combination of atomic orbital (LCAO) type 
formulations, this would be the angular momentum labels $(\ell, m)$  : 

\begin{eqnarray}
H_{\rm lead} &= &\sum_{i,\mu}\sum_{j,\mu'} \hat{H}^{\mu\mu'}_{ij} \vert j,\mu'\rangle\langle i,\mu\vert \nonumber\\
\hat{H}_{ij}^{\mu\mu'}&=&\epsilon^{i\mu}_{lead}\delta_{\mu\mu'}\delta_{ij} + t^{i\mu,j\mu'}_{lead} \delta_{j,i+\chi}
\label{lead}
\end{eqnarray}

 \n where $\chi$ are the $N_i$  nearest neighbors of site $i$ on the lattice. To obtain the equivalent
 `chain' or a tri-diagonal representation of the  Hamiltonian, one needs to change the tight-binding basis $\lbrace|i\mu\rangle\rbrace$ to
 a new one : $\lbrace |n\ggg\rbrace$, obtained recursively.
 
Taking $\vert 1\ggg  = \vert 1\mu\rangle$ as the starting state, where 1 labels the site  at the middle of
the cross-sectional edge of the lead where it is connected to the wire (cf. Fig.1a) and
$\mu$ is any one of the `orbital' indices, we generate :

\begin{equation}
|n+1\ggg  =  H_{\rm lead} |n\ggg - \alpha_n^\mu |n\ggg - {\beta_n^\mu}^2 |n-1\ggg
\end{equation}

\noindent and 

\begin{equation}
\alpha_n^\mu = \frac{\lll n\vert H_{\rm lead}\vert n\ggg}{\lll n\vert n \ggg}\enskip
\beta_n^\mu = \frac{\lll n-1\vert H_{\rm lead}\vert n\ggg}{\left[\lll n\vert n\ggg\lll n-1\vert n-1\ggg\right]^{1/2}}  
\end{equation}

 \n  The equivalent Hamiltonian in this new basis is tri-diagonal (chain-like) : 

 \begin {equation}
\bar{H}_{\rm lead}^\mu =\sum_{n=1}^{\infty}\alpha_n^\mu |n\ggg\lll n| + \beta_n^\mu (|n\ggg\lll n+1| + |n+1\ggg\lll n|)
\end{equation}

\n  where the index $n$ labels the `atoms'  of the equivalent linear chain for the $\mu$ channel and 
$\alpha_n^\mu$ and $\beta_n^\mu$ signify its on-site and the hopping terms respectively. 
The sequences $\{\alpha_n^\mu,\beta_n^\mu\}$ converge, so that for a given error tolerance $\varepsilon$,
there exists an integer $c$ such that for $n>c,\ \vert \alpha_n^\mu-\alpha_c^\mu\vert < \varepsilon$ and
$\vert \beta_n^\mu-\beta_c^\mu\vert < \varepsilon$. The `terminator' approximation puts $\alpha_n^\mu,\beta_n^\mu = \alpha_c^\mu,\beta_c^\mu$ for all $n>c$. 
The initial $c$  `sites'  of both the equivalent input and output chain leads, therefore, contribute to the effective scattering 
region in addition to the wire. The wire of length 2$M$ atoms is described by 
a Hamiltonian in the tight-binding basis as:

\begin {eqnarray}  
 H_{\rm wire}&=& \sum_{i,\mu}\sum_{j,\mu'} \widetilde{H}^{\mu\mu'}_{ij}|j,\mu'\rangle\langle i,\mu|\nonumber \\
\widetilde{H}^{\mu\mu'}_{ij}&=&\epsilon^{i\mu}_{\rm wire}\delta_{\mu\mu'}\delta_{ij} + t^{i\mu,j\mu'}_{\rm wire}(\delta_{j,i+1}+\delta_{j,i-1})
\end{eqnarray}

 The opposite ends of the wire are coupled to the 
 semi-infinite effective 1D leads via hopping matrix element $t_c$. Note that as the starting state is same in the recursion 
process, the coupling coefficient $t_c$ remains same during the change of tight-binding basis set. The procedure therefore converts the whole system into an
 infinite linear chain in which effective scattering region is of extension $2M + 2c$ sites and the rest represents the ballistic parts of the leads which do not participate in the scattering process. 

\subsection{\label{sec:step2}Calculation of the scattering matrix}

\begin{figure}
\begin{center}
\resizebox{0.44\textwidth}{!}{%
\includegraphics{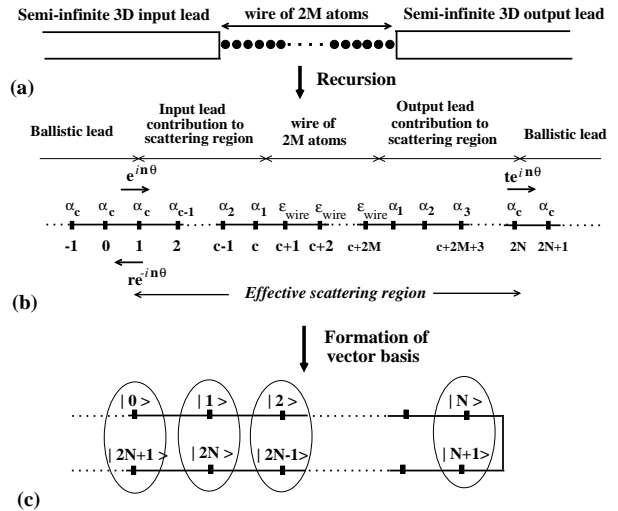}
}
\caption{Recursive reduction of (a) a system of two quasi 1D semi-infinite leads plus 1D wire
into (b) a system of infinite linear chain. The numbering of sites of the equivalent 1D
 infinite chain is shown in (b). The directions of incident wave $e^{in\theta}$ to the scattering
 region, the reflected wave $r e^{-in\theta}$ and transmitted wave $t e^{in\theta}$ from the 
scattering region are shown with arrows. (c) Formation of vector basis by folding the infinite
 chain and clubbing the two sites together.}
\label{fig:recursion}
\end{center}
\end{figure}

Let us consider $2M+2c=2N$ and for convenience, rename the sites 
as follows (cf. Fig.\ref{fig:recursion}b) : 

\begin{eqnarray}
H\phantom{xx} & = & \sum_{n,\mu}\sum_{n',\mu} \widehat{H}^{\mu\mu'}_{nn'} \vert n'\mu'\rangle\langle n\mu\vert \nonumber\\
\widehat{H}^{\mu\mu'}_{nn'} & = & \tilde{\epsilon}^\mu_n \delta_{\mu\mu'}\delta_{nn'} + v^{\mu\mu'}_n \left(\rule{0mm}{4mm}\delta_{n',n+1}+\delta_{n',n-1}\right)
\end{eqnarray}

with 

\begin{equation} 
\tilde{\epsilon}_n^\mu  = \left\{ \begin{array}{lll}
			\alpha_c^\mu & n\le n_1 & \mbox{\ left ballistic lead}\\
                      \alpha_{c+1-n}^\mu & n_1 < n \le n_2& \mbox{\ left scattering lead} \\
                      \epsilon^\mu_{\rm wire} & n_2 < n \le n_3& \mbox{\ scattering wire}\\
                      \alpha^\mu_{n-(c+2M)} & n_3 < n \le n_4 & \mbox{\ right scattering lead} \\
                \alpha^\mu_c &  n > n_4 & \mbox{\ right ballistic lead}
                       \end{array}\right.
\nonumber\end{equation} 

and

\begin{equation}
v^{\mu\mu'}_n =  \left\{ \begin{array}{lll}
             \beta^\mu_c & n \le n_1 & \mbox{\ left ballistic lead}\\
              \beta^\mu_{c-n} & n_1 < n < n_2 & \mbox{\ left scattering lead}\\
               t^{\mu\mu'}_c & n = n_2 & \mbox{\ left junction}\\
              t^{\mu\mu'}_{\rm wire} & n_2 < n < n_3 & \mbox{\ scattering wire} \\
               t^{\mu\mu'}_c &  n = n_3 & \mbox{\ right junction}\\
            \beta^\mu_{n-(c+2M)} & n_3 < n \le n_4 & \mbox{\ right scattering lead}\\
             \beta^\mu_c &  n > n_4 & \mbox{\ right ballistic lead}
	\end{array}\right.
\nonumber\end{equation}
 
\n where $n_1=0,\ n_2=c,\ n_3=c+2M$ and $n_4 = 2c+2M=2N$.

 We then compute the scattering matrix of the above mentioned system using the block or  vector recursion technique introduced by Godin and Haydock  \cite{26}. The
 essence of the vector recursion technique is the block tridiagonalization of the system Hamiltonian by
 changing to a new orthogonal set of vector basis, with the restriction that the ballistic part of effective
 1D lead Hamiltonian 
 remains unchanged. In this last aspect, it differs from the standard Lanczos method \cite{27}. The numerical
 stability of this method \cite{30} has been established in studying problems related to Anderson localization 
 and quantum percolation model \cite{31} and layering transition in 2D nano-strip \cite{29} previously.
Below we describe the method briefly.

Let us consider, for the sake of demonstration, that we have two 1D ballistic leads, one incoming and and another 
outgoing connected to the opposite ends of the scattering region at positions $|1\rangle$ and $|2N\rangle$ (cf. Fig.\ref{fig:recursion}b). The leads and the scatterer have $\mu=1,2,\ldots L$ scattering channels. The 
member of the new basis are generated by clubbing the input and output leads together as the vector lead and
 so is the scattering region (cf. Fig.\ref{fig:recursion}c). 
 The lead states are chosen to be 
\begin{eqnarray}
|\Phi_n\rbrace = \left(\begin{array}{c}|n,\mu=1\rangle\ \ |n,\mu=2\rangle \ldots |n,\mu=L\rangle \\|m,\mu=1\rangle\ \ |m,\mu=1\rangle\ldots  |m,\mu=L\rangle \end{array}\right)^\dagger
\end{eqnarray}

with $m = 2N+1-n$ and $n$ = $0, -1, -2, \dots, \infty$. The starting state within the scattering region is chosen to be
\begin{eqnarray*}
|\Phi_1\rbrace = \left(\begin{array}{c}|1,\mu=1\rangle\ \ |1,\mu=2\rangle \ldots |1,\mu=L\rangle
\\ |2N,\mu=1\rangle\ \ |2N,\mu=2\rangle \ldots |2N,\mu=L\rangle\end{array}\right)^\dagger
\end{eqnarray*}

 The subsequent members of the basis are generated from 
\begin{eqnarray}
B_2^\dagger |\Phi_2\rbrace = ( H - A_1 ) |\Phi_1\rbrace,\phantom{xxxxxxxxxxxxxxxxxxxx}\nonumber\\
B_{n+1}^\dagger |\Phi_{n+1}\rbrace = ( H - A_n)|\Phi_n\rbrace - B_n|\Phi_{n-1}\rbrace \phantom{x}  \mbox{for \phantom{x}} n \geq 2\nonumber \\
\label{eqn}
\end{eqnarray}

 The matrix inner product is defined as the $2L\times 2L$ matrix 

\[ \{\Phi_1|\odot|\Phi_1\}  = \left( \begin{array}{cc} A & B \\ C & D \end{array}\right)
\]

\n where the $L\times L$ matrices $A,B,C$ and $D$ are :
 
\begin{eqnarray*}
A&=& \langle 1\mu | 1\mu'\rangle \qquad 
\phantom{B} B= \langle 1\mu | 2N\mu'\rangle \\ 
C&=& \langle 2N\mu | 1\mu'\rangle \qquad 
D= \langle 2N\mu | 2N\mu'\rangle  
\end{eqnarray*}

If this matrix is $I$ then the states are called orthogonal.

It can be shown that the $2L\times2L$ matrices $A_n$ and $B_n$ are block-tridiagonal members of the matrix representation
 of the Hamiltonian in the new basis:
\begin{eqnarray}
A_n = \lbrace\Phi_n|\odot H|\Phi_n\rbrace \qquad B_n = \lbrace\Phi_n|\odot H|\Phi_{n-1}\rbrace
\end{eqnarray}

so that the transformed Hamiltonian matrix can be divided in $2L\times2L$ blocks, with only non-zero diagonal and 
sub-diagonal blocks.

The wave function $|\Psi\rbrace$ may be represented in this new basis by 
 a set $\lbrace\psi_n\rbrace$ so that 
$|\Psi\rbrace$ = $\sum_n\psi_n|\Phi_n\rbrace$. These wave function amplitudes $\psi_n$ also satisfy an equation 
identical with (\ref{eqn}).\\

In the ballistic part of the effective 1D lead chain, the onsite and hopping terms do not vary. Therefore, 
electron potential must be periodic in this region and the solution to the Schr\"odinger's equation 
in the $\mu$ channel of the ballistic leads  are traveling Bloch waves : 
$A\ \sum_m  \exp[\pm im\theta^\mu]|m\rangle$.
As the wave travels in the leads, the phase of its wave function changes by $\theta^\mu = 
\cos^{-1}\left[(E - \alpha^\mu_c)/(2 \beta^\mu_c)\right]$, where $E$ is the energy of the incoming electron. In order to have only propagating solutions, we fix E as real and $| E -\alpha_c | < 2\beta_c$. This sets the energy window. This is reasonable because eventually only the propagating modes enter in the expression of transmission matrix elements, as evanescent modes do not contribute to the transmission directly. In the $\mu$ channel of the incoming lead, the incident and the reflected waves can be expressed as a sum 

\[A\sum_m \sum_{\mu'}\left[\rule{0mm}{3mm} \exp(im\theta^\mu)\delta_{\mu\mu '}+r^{\mu\mu'}(E)\ \exp(-im\theta^{\mu'})\right]|m\rangle\]

The second term is the reflected wave in the $\mu$ channel from incident waves in $\mu '$ channels.

 In the $\mu$ channel of the output lead there is a transmitted wave from incident waves in $\mu '$ channels
 \cite{33} 

 \[ A \sum_m\sum_{\mu'} t^{\mu\mu'}(E)\ \exp(-im\theta^{\mu'})|m\rangle \]  

 $r^{\mu\mu'}(E)$ and $t^{\mu\mu'}(E)$ are the complex reflection and transmission coefficients. The boundary conditions 
are  then  imposed from the known solution in the leads at the junction labeled by 0 and 1:
\begin{eqnarray}
\psi_0 = \left(\begin{array}{c}\sum_{\mu'}[\delta_{\mu\mu'} + r^{\mu\mu'}(E)]\\\sum_{\mu '}t^{\mu\mu'}(E)\end{array}\right)
\end{eqnarray}

\begin{eqnarray}
\psi_1 = \left(\begin{array}{c}\sum_{\mu '}\left\{\rule{0mm}{3mm}\exp[i\theta^\mu]\delta_{\mu\mu'} + r^{\mu\mu'}(E) \exp[-i\theta^{\mu'}]\right\}\\\sum_{\mu '}t^{\mu\mu'}(E) \exp[-i\theta^{\mu\mu'}]\end{array}\right)
\end{eqnarray}

The amplitude at the $n$-th basis $\psi_n$ may be written as 
\begin{eqnarray}
\psi_n = X_n\psi_0 + Y_n\psi_1
\end{eqnarray}

where $X_n$ and $Y_n$ satisfy the same recurrence relation as (\ref{eqn}) with $EI$ replacing $H$ and
 also satisfy the boundary conditions $X_0$ = $I$ and $X_1$ = $0$, while $Y_0$ = $0$ and $Y_1$ = $I$. Note
 that $X$ and $Y$ are $2L\times2L$ matrices and $\psi_n$'s are column matrices of dimension 2$L$.

This new basis terminates after $\nu$ = $N$ steps, as the rank of the space spanned by the
 original tight-binding basis remains unchanged after the transformation. Hence the recursion also terminates 
 after $\nu$ steps. This gives an additional boundary condition,
\begin{eqnarray}
X_{\nu+1}\psi_0 + Y_{\nu+1} \psi_1 = 0_{2L\times2L}.
\end{eqnarray}

If we now interchange the incoming and outgoing leads, we get a similar pair of equations
 for ${r^\prime}^{\mu\mu'}$ and ${t^\prime}^{\mu\mu'}$, the transmission and reflection coefficients for a wave incident from the second
lead. Time-reversal symmetry demands that $t$ must be the same for waves of the same energy incident from 
either lead so that $t^{\mu\mu'}$ = ${t^\prime}^{\mu\mu'}. $ 
 In addition,  $r^{\mu\mu'}$
 and  ${r^\prime}^{\mu\mu'}$ differ only by a phase factor. Solving these equations for the S-matrix in the effective scattering region,
 one has,
\begin{eqnarray}
S(E) &=& -(X_{N+1} + Y_{N+1} E^*(\theta))^{-1}(X_{N+1} + Y_{N+1} E(\theta)) \nonumber\\
{\hskip -0.5cm} &=& \left(\begin{array}{cc}r^{\mu\mu'}(E) & t^{\mu\mu'}(E) \\t^{\mu\mu'}(E) & {r^\prime(E)}^{\mu\mu'}\end{array}\right)
\end{eqnarray}

\n where

\[ E(\theta) = \left(\begin{array}{cc}  \exp(i\theta^\mu)\delta_{\mu\mu'} & 0 \\
                       0 & \exp(i\theta^\mu)\delta_{\mu\mu'}\end{array}\right) \]
The various steps of our approach are shown schematically in Fig.~\ref{fig:recursion}.

The conductance is then given by Landauer formalism. It expresses the electronic conductance
in one-dimensional conductor as a quantum mechanical scattering problem and relates
to the total transmission probability of the electron at the Fermi level\cite{32}, T($E_F$), as

\begin{equation}
\label{lbeq}
 G=\left(\frac{e^2}{\pi \hbar}\right)T(E_F) = \left(\frac{e^2}{\pi \hbar}\right) \left\vert (1/L)\sum_{\mu\mu'} t^{\mu\mu'}(E_F)\right\vert^2
\end{equation}

We have assumed in our derivation the leads to have finite lateral dimensions, though in actual experiments, 
the leads are three dimensional. In order to study the influence of the extend of the lateral dimension of
 the leads, we have checked our results with increasing size of the lateral cross-section of the leads. Though
 the quantitative values do change with the change of the lateral dimension, the qualitative nature holds good
in every case. In addition, we find for sufficiently large choice of lateral
 cross-section, the recursion coefficients ($\alpha$, $\beta$) for the leads converge to that of a bulk
 system (see for details section 3.1),
 indicating for such large lateral cross-section, the leads behave as truly three-dimensional leads.
 The other approach to the problem could have been inclusion of periodic boundary condition as have been adopted in various works \cite{lcao,wavelet,wannier,10,pbc1,pbc2,pbc3}. If we impose periodic boundary conditions on the lead surfaces, only those modes which are consistent
with the boundary conditions can travel through them. In an earlier paper we have shown \cite{mtm}, that
it is possible to change over from a site to a mode basis and reformulate the vector recursion in the new basis.
The lead is broken up into slices perpendicular to its length and the composite site label $i$ is
partitioned into two : one is the slice label $s$ and the other is the position on this slice $k$. The
basis $\{|sk\rangle\}$ is then converted into a slice-mode basis $\{ |s\nu\rangle\}$ and the
Hamiltonian is expressed in this new basis.  Exactly as the Hamiltonian (6) mixes
orbital labeled channels because of off-diagonal terms $v^{\mu\mu'}$, in the mode based formalism
off-diagonal terms in the corresponding Hamiltonian causes the outgoing wave to be of a mixed mode
type even if the incoming wave is in a single mode. Otherwise, the formal methodology is identical in the
two cases. The interested reader is referred to the paper referenced above for the details.

\section{\label{sec:application} Application}

\subsection{\label{sec:model} The model }

 \begin{figure}
\resizebox{0.44\textwidth}{!}{%
\includegraphics{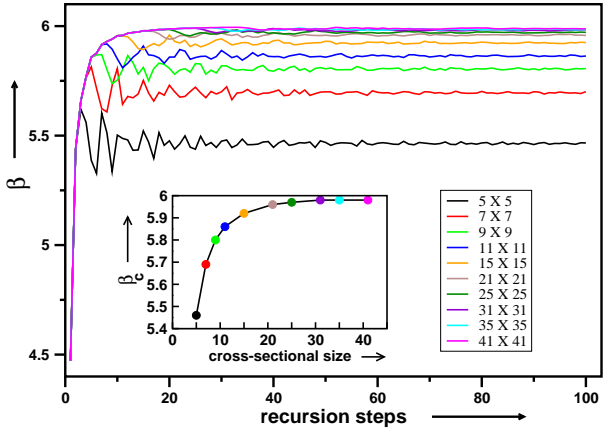}
}
\caption{(Color online) Variation of recursion coefficient $\beta$ with recursion steps for different cross-sectional 
sizes of the lead, with t$_{lead}$ = 2.0 and $\epsilon_{lead}$ = 0.0. The lower left inset shows the 
variation of converged values of $\beta$ ($\beta_c$) with cross-sectional size.}
\label{fig:beta}
\end{figure}
\begin{figure}
\begin{center}
\resizebox{0.49\textwidth}{!}{%
  \includegraphics{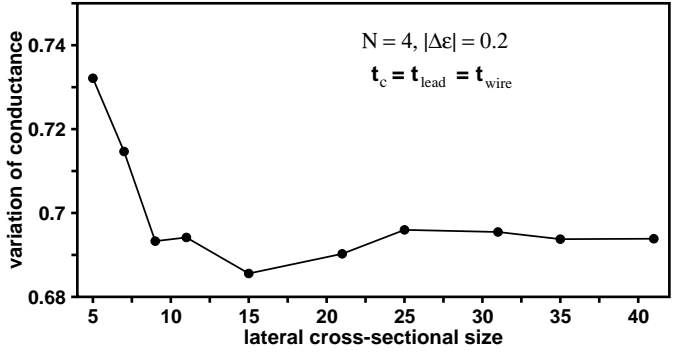}
}

\caption{Variation of conductance with cross-sectional sizes for an even numbered wire, $N$ = 4
considering direct coupling and $|\Delta\epsilon|$ = 0.2.
 Variation in conductance is around 5\% for even numbered wire. For odd numbered wire (not shown in the
above plot), it is less than 1\%.}
\label{fig:convg_cond}
\end{center}
\end{figure}

Since our aim here is to propose a method rather than to evaluate the properties of any particular 
system in quantitative detail, we present the study of a  simple model system.  
 It consists of a wire,  few atoms long, sandwiched 
between two identical semi-infinite leads. Both the wire and the leads have only a single
channel ($L = 1)$ corresponding to single $s$-band. The cross-sectional size of each lead in our calculation
  was 5$\times$5. The atoms in the leads form a simple cubic lattice. Each lead is described by
 a single-band nearest-neighbor TB Hamiltonian. Two TB parameters in equation (\ref{lead}) : the on-site  
and the hopping terms, which are material specific in realistic cases, have been chosen as 
$\epsilon^i$ = $\epsilon_{\rm lead}$ = 0 and $t^{ij}$ = $t_{\rm lead}$ = 2 in some arbitrary energy unit. As a first step, we performed  scalar recursion on the leads to convert them  into equivalent chains and determined the 
contribution of each lead to the
 effective scattering region. We took the starting state of recursion to be the sites which bound the leads 
 to the wire. For a simple cubic lead, the recursion coefficients  
$\alpha_n = \epsilon_{\rm lead}$ = 0 while the other coefficients $\beta_n$ 
fluctuate with recursion steps converging to an asymptotic value $\beta_c$. The 
convergence of $\beta_n$ depends on the lattice structure and lateral cross-sectional size of lead. We have not used
 any periodic boundary conditions in the lateral direction. To see the finite size effect of lead in the lateral
 direction, we have repeated our calculations for different cross-sectional sizes. Convergence of $\beta$ with 
recursion steps for different cross-sectional sizes is shown in Fig.\ref{fig:beta}. As the cross-section increases,
 the number of recursion steps required to converge $\beta$ decreases (i.e size of effective scattering region
 decreases which facilitates to easier computational execution). Variation of converged value $\beta$ ($\beta_c$) 
with lateral cross-sectional size of the lead is shown within the inset. Notice after a certain cross-sectional 
size (30 $\times$ 30 with the present choice of parameters), $\beta$ converges to that of the bulk cubic system, which is 6 for choice of $\epsilon_{\rm lead}$ = 0 and  $t_{\rm lead}$ = 2.0. This implies that there is no further effect of finite cross-section on $\beta_c$ and therefore on 
conductance. In changing cross-section from 5$\times$5 to 30$\times$30, $\beta_c$ changes by 9 \% (The change
 is mostly coming in going from 5$\times$5 to 11$\times$11 and after that change is very minute). The 
corresponding change in conductance is much smaller : 5 \% for even numbered wire, less than 1 
\% for odd numbered wire (cf. Fig. \ref{fig:convg_cond}). We have checked that qualitatively results are
 same for any cross-sectional size, but if one is concerned about the quantitative value, then convergence of 
$\beta_c$ with cross-sectional size has to be checked. For our chosen case, $\beta_n$ converged after
 around 80 recursions. So in our calculation, $c$ was taken to be 80 and $\beta_n =\beta_{80}$ for $n\geq 80$. For a model simple cubic lead, the Fermi energy 
is at $E_F^{\rm lead}$ = $\epsilon_{\rm lead}$ which is obvious from the dispersion relation $E = \epsilon_{\rm lead} 
+ 6 t_{\rm lead} \cos(ka)$ for a half-filled band with Fermi wave number $k_F$ = $\pi / 2a$. Fermi energy $E_F$ of the lead-wire-lead system tends to align with the Fermi
 energy of the semi-infinite lead. Conductance is therefore calculated  at $E_F$ = $\epsilon_{\rm lead}$.

\subsection{\label{sec:result} Results and discussions}

\begin{figure}
\begin{center} 
\resizebox{0.51\textwidth}{!}{%
  \includegraphics{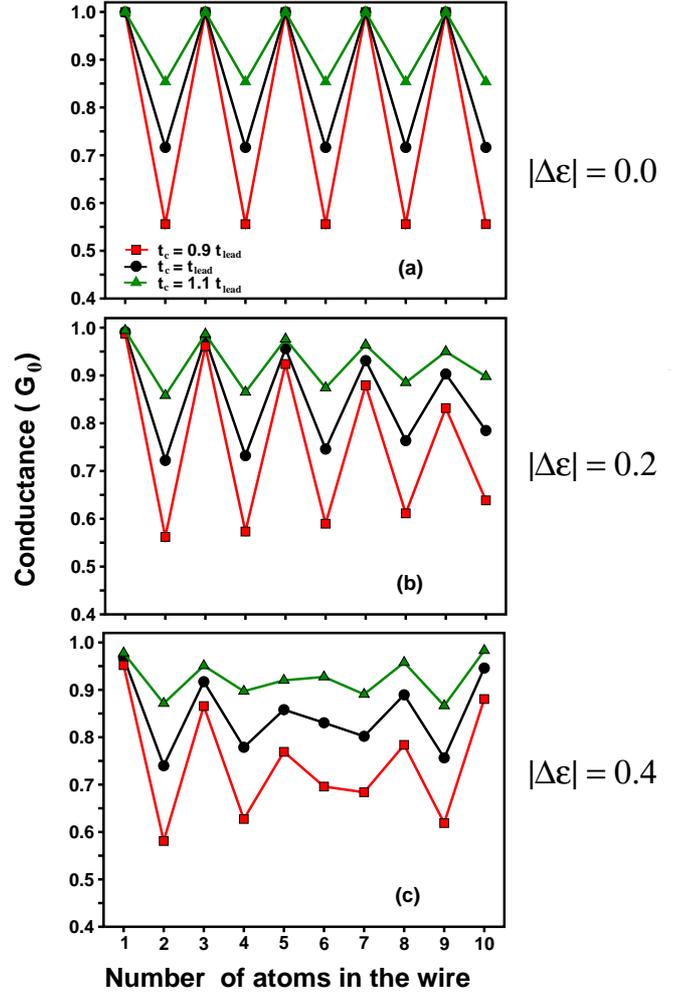}
}

\caption{(Color online) Conductance as a function of the number of atoms in the wire. For given values
 of $t_{\rm lead}$ and $t_{\rm wire}$, we consider three cases: (a) $|\Delta\epsilon|$ = 0,
 (b) $|\Delta\epsilon|$ = 0.2 and (c) $|\Delta\epsilon|$ = 0.4. For each case, we consider three 
types of coupling between lead and wire : green curve for strong coupling, black for direct
 coupling and red curve for weak coupling. }
\label{fig:conductance}
\end{center}
\end{figure}

Our calculated conductance as a function of number of atoms in the wire is shown in 
Fig.\ref{fig:conductance}. 
 The inter-atomic hopping 
 within the wire was fixed to $t_{\rm wire}$ = 2 = $t_{\rm lead}$ . To consider the effect of charge transfer
 between the leads and 
the wire [$\Delta \epsilon$ = $\epsilon_{\rm lead}$ - $\epsilon_{\rm wire}$ $\not=$ 0], we considered 
 the cases $|\Delta\epsilon|$ = 0.2 and $|\Delta\epsilon|$ = 0.4 along with no charge transfer 
condition  $|\Delta\epsilon|$ = 0. Non zero values of $\Delta \epsilon$ were achieved by changing $\epsilon_{\rm wire}$. 
For each case, we considered three types of coupling between lead and atomic wire, given by 
lead-wire hopping coefficient $t_c$ : (i) $t_c$ = 0.9 $t_{lead}$ for weak coupling, (ii) $t_c$ 
= $t_{lead}$ for direct coupling and (iii) $t_c$ = 1.1 $t_{lead}$ for strong coupling.

The odd-even oscillation in conductance for no charge transfer situation is obvious from Fig.\ref{fig:conductance}a {\it i.e}  wires with odd-number of atoms have larger conductance as
compared to  
the even numbered ones, for all the three types of couplings. In this ideal case of no 
charge transfer situation, all odd numbered atomic wires have conductances equal to 
quantum unit $G_0$ = $(e^2/\pi \hbar)$ for all the three coupling cases, while for even numbered wires the conductances
 are lower. The conductance of even-numbered wires decreases as one moves
 from strong to direct to weak
 coupling. This means that the amplitude of conductance oscillation is a maximum for the weak coupling 
case ($t_c$ = 0.9 $t_{lead}$) and it is a minimum for the strong coupling case ($t_c$ = 1.1 $t_{lead}$).
Similar results have been predicted by Khomyakov {\it et al} for a system of 1D wire connected with 1D leads 
using tight-binding calculation \cite{8}. The dependence of conductance on $t_c$ for even numbered
 wire is given by 
\[ G = G_0\frac{4t_c^4/t_{\rm wire}^2}{[1 + 4t_c^4/t_{\rm wire}^2]^2}\]

It follows the same trend as observed
 in the top panel of Fig.\ref{fig:conductance}. Another point to notice 
from Fig.\ref{fig:conductance}a is that the length of the wire does not have any effect in this case 
either on amplitude or  phase of the oscillation. All odd numbered atomic wires  and also all 
 even numbered atomic wires have equal conductances for a fixed coupling type. The 
odd-even oscillations for our single band model system is in agreement with the results
on Na-atom wires\cite{3,5,6,7,8,ex6,ex2}.

 For $|\Delta\epsilon|$ = 0.2 in Fig.\ref{fig:conductance}b, the odd-even oscillation in 
conductance is again obvious within the range of our plot. However, a close view to these curves,
 indicates that the conductance of the odd numbered atomic wires gradually decreases while that of the  even 
numbered atomic wires slowly increases. If we increase the number of atoms in the wire to more 
than 10 (not shown in Fig.\ref{fig:conductance}), we find that the parity of conductance
 oscillation changes from odd-even to even-odd, {\it i.e} even numbered atomic wires now have larger 
conductance than the conductance of odd numbered atomic wires.

 Increasing $|\Delta\epsilon|$ to 0.4,  there will be  larger   charge transfer 
between the lead and wire. This causes a change in the parity of conductance 
oscillation from odd-even to even-odd for even shorter lengths :  around wire lengths
  of 7 atoms as shown in
 Fig.\ref{fig:conductance}c. From our study, it is therefore clear that the amplitude of conductance
 oscillation is mainly controlled by coupling coefficient between the leads and the wire, while the parity of 
the oscillation is controlled by the length of the wire and onsite energy difference between wire
 and lead Hamiltonians. There are few reports \cite{ex6,ex5} which indicate the change of
 parity of conductance oscillation and variation of oscillation amplitude for Na-atom wire.

 \begin{figure}
\begin{center}
\resizebox{0.48\textwidth}{!}{%
  \includegraphics{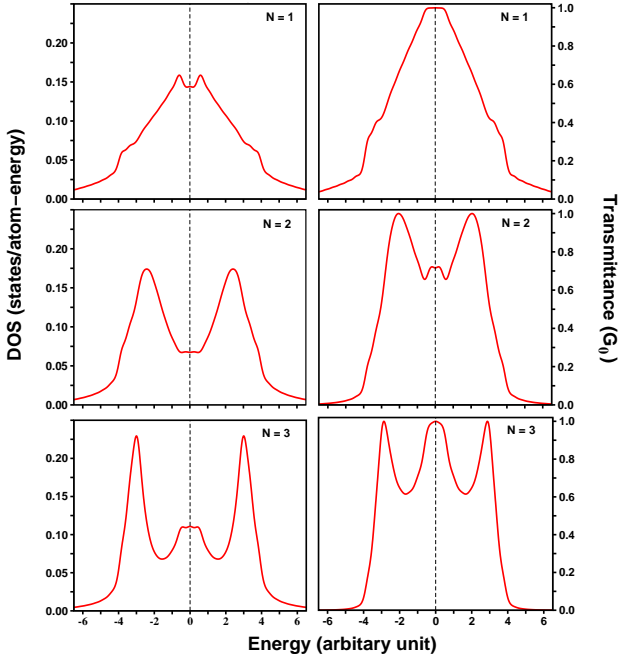}
}

\caption{(Color online) Density of states (left panel) and total transmittance (right panel) in
 no charge transfer ($|\Delta\epsilon|$ = 0) and direct coupling ($t_c$ = $t_{lead}$) situation
 for wires of lengths $N$ = 1, $N$ = 2 and $N$ = 3 (from top to bottom). The dashed lines
label the Fermi energy.  }
\label{fig:dos}
\end{center}
\end{figure}
To understand the odd-even oscillation in the conductance, we have calculated the 
density of states (DOS) and transmittance as a function of energy for wires of lengths
 $N$ = 1, $N$ = 2 and $N$ = 3 for the case of zero charge transfer. The results shown in 
Fig.\ref{fig:dos} bring out the essential
 mechanism which has been discussed also in ref. \cite{8}. For odd-numbered atomic wires, the Fermi-energy
 falls in the non-bonding peak and it gives a maximum in the transmittance at $E$ = $E_F$. On the 
 other hand, for even numbered wire, the Fermi level lies in the minimum between the bonding and non-bonding
 peaks and consequently it exhibits a less transmission at $E$ = $E_F$.

 \begin{figure}
\begin{center}
\resizebox{0.48\textwidth}{!}{%
  \includegraphics{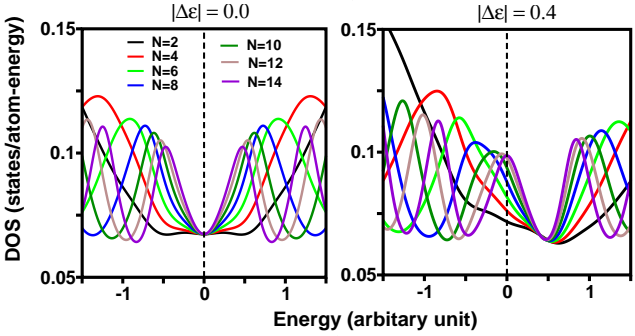}
}

\caption{(Color online) DOS around Fermi energy for several even numbered wires in two cases - $|\Delta\epsilon|$ = 0
 (left) and $|\Delta\epsilon|$ = 0.4 (right). For $|\Delta\epsilon|$ = 0.4, DOS at $E_F$ gradually 
increases with wire size. For $N$ = 14, the deep at $E_F$ for $|\Delta\epsilon|$ = 0 case is 
replaced by a peak at $E_F$ for $|\Delta\epsilon|$ = 0.4. Lead-wire coupling
 was considered to be of direct type in all cases. }
\label{fig:dos_even}
\end{center}
\end{figure}

In order to explain the flipping of conductance oscillation from odd-even to even-odd, we have
plotted  in Fig.\ref{fig:dos_even} the DOS around Fermi energy of several even numbered atomic 
wires for two cases of no charge transfer and finite charge transfer situations. When there is
 no charge transfer ($|\Delta\epsilon|$ = 0), all the even 
numbered wires have a minimum in DOS at $E_F$. When we allow sufficient charge transfer to take place 
between wire and lead, DOS at $E_F$ gradually increases with increasing length and at a critical size, 
a peak will appear in DOS at $E_F$. The opposite  occurs for odd-numbered atomic wire (not shown 
in the Fig.\ref{fig:dos_even}) and nature of conductance oscillations flips from odd-even to even-odd.
 If we increase the wire size further, again after another critical size, the odd-even nature of 
oscillations is restored. This repetition of even-odd or odd-even oscillation continues with increasing
 wire size.

 \begin{figure}
\begin{center}
\resizebox{0.48\textwidth}{!}{%
  \includegraphics{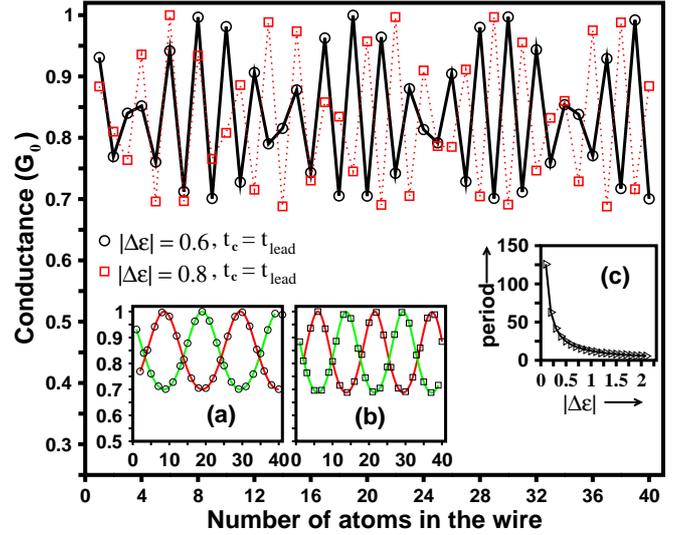}
}

\caption{(Color online) Plot of conductance as a function of wire size for two values of
 $|\Delta\epsilon|$ : circles connected by solid line for $|\Delta\epsilon|$ = 0.6 and squares 
connected by dashed line for $|\Delta\epsilon|$ = 0.8. Conductances for all even numbered 
wires and all odd numbered wires are shown separately by red line and green line respectively within
 the insets. Inset (a) for $|\Delta\epsilon|$ = 0.6 and inset (b) for $|\Delta\epsilon|$ = 0.8. Inset (c)
 shows the variation of period of parity flip with $|\Delta\epsilon|$ and solid line through the right 
triangles is the fitted curve. }
\label{fig:cond_ewire_vary}
\end{center}
\end{figure}

To investigate the effect of $|\Delta\epsilon|$ on the period of parity flip, we have plotted
 the conductance as a function of number of atoms in the wires in Fig.\ref{fig:cond_ewire_vary}
 for two different values of $|\Delta\epsilon|$ keeping coupling parameter fixed. One can see clearly 
that as $|\Delta\epsilon|$ increases, frequency of parity flip increases {\it i.e} period of parity flip
 decreases. To get a better insight into  parity flip, we have shown in 
Fig.\ref{fig:cond_ewire_vary} the conductance variation of odd numbered and even numbered wires 
separately in insets (a) and (b) for $|\Delta\epsilon|$ = 0.6 and $|\Delta\epsilon|$ = 0.8 respectively.
 Each curve shows sinusoidal-like variations. Curves of even numbered and odd numbered wires together
 constitute loops. Length of one loop indicates the wire size required to flip the oscillation 
from odd-even to even-odd. The nodal points of loop indicate the boundaries between odd-even 
and even-odd. Two consecutive loops
 constitute a period. Number of loops in inset (b) is larger than in inset (a). Variation of period
of parity-flip with $|\Delta\epsilon|$ is shown by right triangles in inset (c). For $|\Delta\epsilon|$ = 0,
 there is no parity flip {\it i.e} period is infinity. As $|\Delta\epsilon|$ increases, the period
 decreases. For sufficiently large value of $|\Delta\epsilon|$ ($>> t_c$), odd-even nature of 
conductance oscillation no longer persists. Period of conductance oscillation then changes to more 
than two atoms which is the characteristic of wires consisting of atoms of higher valency. Within the
 odd-even nature of conductance oscillations, the period of parity flip goes roughly as 
$(A/|\Delta\epsilon|) + B$ with $A$ = 12.605 and $B$ = -0.253. 

 \begin{figure}
\begin{center}
\resizebox{0.48\textwidth}{!}{%
  \includegraphics{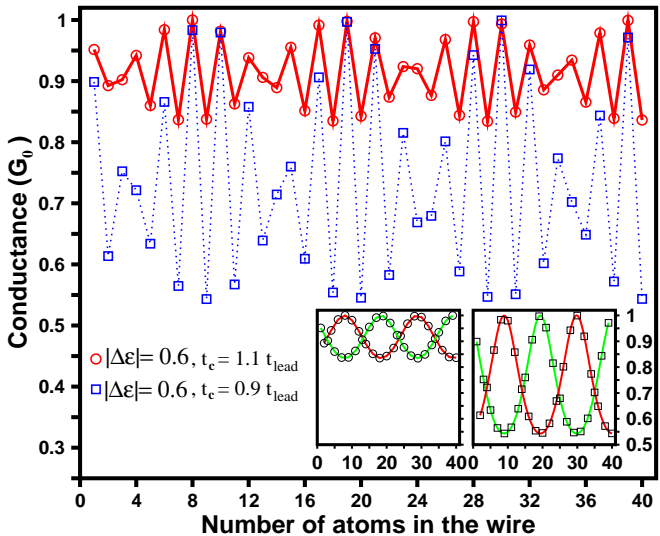}
}

\caption{(Color online) Plot of conductance as a function of wire size for two values of $t_c$ - 
circles connected by solid line for $t_c$ = 1.1 $t_{lead}$ and  squares connected by dashed line for $t_c$ = 0.9 $t_{lead}$.
Conductances for all even numbered wires and all odd numbered wires are shown separately by red line 
and green line respectively. Left inset corresponds to $t_c$ = 1.1 $t_{lead}$ and right one to $t_c$ = 0.9 $t_{lead}$ .  }
\label{fig:cond_tc_vary}
\end{center}
\end{figure}
However, coupling parameter $t_c$ does not have any effect on period of parity flip. To check this,
 we have plotted the conductance variation with wire size in Fig.\ref{fig:cond_tc_vary} for two 
different coupling constant $t_c$ keeping $|\Delta\epsilon|$ fixed, while insets (a) and (b) show the
 conductance variation of odd numbered wires and even numbered wires separately. The number of loops 
in both the cases are same indicating $t_c$ has no effect on the period. Larger width of loops
 in inset (b) compared to that in inset (a) indicates that $t_c$ controls the amplitude of odd-even oscillations .

\begin{table}
\caption{Values of conductances for various wire lengths in absence of mirror symmetry.
 $t_{c1}$ and $t_{c2}$ are the two lead-wire couplings of the two junctions.}
\label{tab:1}       

\begin{tabular}{llll}
\hline\noalign{\smallskip}
wire length & t$_{c1}$ = t$_{lead}$  & t$_{c1}$ = t$_{lead}$     & t$_{c1}$ = t$_{lead}$ \\
  (N)       & t$_{c2}$ = t$_{lead}$  & t$_{c2}$ = 1.1 t$_{lead}$ & t$_{c2}$ = 1.2 t$_{lead}$ \\
\noalign{\smallskip}\hline\noalign{\smallskip}
1 & 0.9996  & 0.9906  & 0.9672  \\
2 & 0.7164  & 0.7876  & 0.8482  \\
3 & 0.9996  & 0.9906  & 0.9672  \\
4 & 0.7164  & 0.7876  & 0.8482  \\
5 & 0.9996  & 0.9906  & 0.9672  \\
\noalign{\smallskip}\hline
\end{tabular}

\end{table}

So far we have studied the role of the charge neutrality on the conductance oscillations of monoatomic
 wires considering the mirror symmetry between the two junctions. We found odd-even oscillation in the
 conductance.
 Moreover, for no charge transfer situation, conductance of odd numbered wires is quantized to $G_0$, while
 for even numbered wires conductance is less than $G_0$. Now we consider the situation where 
mirror symmetry between two lead-wire junctions is broken by using two different coupling coefficients for two junctions, keeping
 charge-neutrality intact. Table \ref{tab:1} contains our result. Clearly, odd-even oscillations still occur,
 but the conductance quantization for odd numbered wires is weakened by reducing the value less than  $G_0$.
 This observation is in accordance with the previous study\cite{6,mirror}.

To conclude, we have used combination of real-space based scalar and vector recursion techniques to study
the transport properties of a lead-wire-lead system. Our study on model system described by single 
band TB Hamiltonian provides a detail understanding of the effect of lead-wire coupling on the conductance of
 monoatomic wire. Working with model system, gives us the freedom of changing the model parameters 
and allows us to work with much longer wires than usually considered in liturature. Odd-even oscillation 
in the conductance with increasing length of wire has been observed in agreement with earlier studies \cite{3,5,6,7,8,ex6,ex2,ex5,ex7}. 
In presence of charge neutrality between the leads and the wire and in presence of perfect mirror symmetry 
between the incoming and outgoing leads, the conductance of odd numbered wires is quantized to 
 $G_0$, while it is less than $G_0$ for even numbered wires. As the charge neutrality is broken,
 the oscillation in conductance still exists, but with the distinction that for a given choice of 
charge transfer ($\Delta\epsilon$) and lead-wire hopping ($t_c$), the conductance values of the odd
 numbered and even numbered wires are no longer fixed quantities as the size of the wires is changed.
For such systems, we further found a change of phase of conductance oscillation from odd-even to 
even-odd with increasing number of atoms in the wire. We found that while the amplitude of oscillation
 depends on lead-wire coupling parameter $t_c$, it is the amount of charge transfer between lead and wire, which
 affects the period of oscillation. Lifting of mirror symmetry between two lead-wire junctions in no 
charge transfer condition is found to reduce the conductance of odd numbered wires below
 the quantized value of $G_0$. The proposed technique can be easily generalized for application 
to realistic cases. To apply this approach to real systems, one needs to generate TB parameters 
of lead and wire via self consistent calculations while multi-orbital effect can be taken
 into account via multi-channel generalization of Landauer-B\"uttiker formula (see equation \ref{lbeq}). The proposed 
real space technique of calculation of conductance coupled with localized Wannier basis generated
 out of self-consistent DFT calculation \cite{wannier1,wannier2} can lead to a viable technique 
for study of quantum transmittance and conductance of nanoscale systems of various geometries
 in general. Work is in progress along this direction.


S.D thanks Council of Scientific and Industrial Research (Government of India) for financial support and TSD acknowledges Department of Science and Technology (Government of India) for support through Swarnajayanti fellowship.


\begin{thebibliography}{}

\bibitem{1}
  A. I. Yanson, G. R. Bollinger, H. E. van der Brom, N. Agrait, J. M. van Ruitenbeek,
 Nature \textbf{395}, {1998} 783

\bibitem{add1}
  H. Ohnishi, Y. Kondo, K. Takayanagi, Nature \textbf{395}, (1998) 780


\bibitem{2}
  R. H. M. Smit, C. Untiedt, G. Rubio-Bollinger, R. C. Segers, J. M. van Ruitenbeek,
 Phys. Rev. Lett. \textbf {91}, (2003) 076805

\bibitem{add2}
W. H. A. Thijssen, D. Marjenburgh, R. H. Bremmer, J. M. van Ruitenbeek, Phys. Rev. Lett. \textbf {96}, (2006) 026806

\bibitem{ex1}
  J. M. Krans, J. M. van Ruitenbeek, V. V. Fisun, I. K. Yanson, L. J. de Jongh,
 Nature (London) \textbf{375}, (1995) 767; A. I. Yanson, I. K. Yanson, 
J. M. van Ruitenbeek, ibid. \textbf{400} (1999) 144

\bibitem{huc1}
P. Sautet, C. Joachim, Phys. Rev. B \textbf {38}, (1988) 12238

\bibitem{huc2}
E. G. Emberly, G. Kirzenow, Phys. Rev. B \textbf {58}, (1988) 10911

\bibitem{trfm1}
K. Hirose, M. Tsukada, Phys. Rev. Lett. \textbf {73}, (1994) 150

\bibitem{trfm2}
K. Hirose, M. Tsukada, Phys. Rev. B \textbf {51}, (1995) 5278

\bibitem{trfm3}
H. J. Choi, J. Ihm, Phys. Rev. B \textbf {59}, (1999) 2267

\bibitem{fdm}
P. A. Khomyakov, G. Brocks, Phys. Rev. B \textbf {70}, (2004) 195402

\bibitem{lse1}
N. D. Lang, Phys. Rev. B \textbf {52}, (1995) 5335; N. D. Lang, Phys. Rev. Lett. \textbf {79}, (1997) 1357

\bibitem{lse2}
M. D. Ventra, S. T. Pantelides, N. D. Lang, Phys. Rev. Lett. \textbf {84}, (2000) 979
 
\bibitem{gs}
Y. Xue, S. Datta, M. A. Ratner, Chem. Phys. \textbf {281}, (2001) 151
 
\bibitem{lcao}
M. Brandbyge, J. L. Mozos, P. Ordejon, J. Taylor, K. Stokbro, Phys. Rev. B \textbf {65}, (2002) 165401
 
\bibitem{wavelet}
K. S. Thygesen, M. V. Bollinger, K. W. Jacobsen, Phys. Rev. B \textbf {67}, (2003) 115404
 
\bibitem{wannier}
A. Calzolari, N. Marzari, I. Souza, M.B. Nardelli, Phys. Rev. B \textbf {69}, (2004) 035108

\bibitem{10}
 P. Khomyakov, G. Brocks, V. Karpan, M. Zwierzycki, P. J. Kelly, Phys. Rev. B \textbf {72}, (2005) 
035450


\bibitem{26a}
  R. Haydock, V. Heine, M. J. Kelly, J. Phys. C :Solid State Phys \textbf {5}, (1972) 2845

\bibitem{27}
R. Haydock, \textit{Solid State Physics}, editor: H. Ehrenreich, F. Sietz, D. Turnbull
(Academic, New York, 1980) volume 35

\bibitem{26}
  T. J. Godin,  R. Haydock, Phys. Rev. B \textbf {38}, (1988) 5237

\bibitem{30}
  T. J. Godin, R. Haydock, Comp. Phys. Comm. \textbf {64}, (1991) 123

\bibitem{31}
  I. Dasgupta, T. Saha, A. Mookerjee, Phys. Rev. B \textbf {47}, (1993) 3097

\bibitem{29}
S. Datta, D. Choidhuri, T. Saha-Dasgupta, S. Sengupta, Eup. Phys. Lett. \textbf{73}, (2006) 765

\bibitem{33} The process of vector recursion converts the lattice into a one-dimensional chain, 
which is then folded to clump two sites of the chain together to define the vector basis set 
. For a chain with folded configuration (see Ref. \cite{26} 
for details) both the reflected and transmitted waves move in the opposite direction
 to that of the  incident wave.

\bibitem{32}
  The Fermi energy of the lead-wire composite system is determined by 
the macroscopic lead Hamiltonian.

\bibitem{pbc1}
M. B. Nardelli, Phys. Rev. B \textbf{60}, (1999) 7828

\bibitem{pbc2}

K. S. Thygesen, K. W. Jacobsen, Chem. Phys.  \textbf{319}, (2005) 111


\bibitem{pbc3}
P. A. Khomyakov and G. Brocks, Phys. Rev. B \textbf{70}, (2004) 195402;
 K. Xia, M. Zwierzycki, M. Talanana and P. J. Kelly, Phys. Rev. B \textbf{73}, (2006) 064420

\bibitem{mtm}
K. Tarafder, T. Mitra and A. Mookerjee, Physica B \textbf{371}, (2006) 100

\bibitem{3}
  N. D. Lang, Phys. Rev. Lett. \textbf{79}, (1997) 1357 

\bibitem{5}
  S. Tsukamoto, K. Hirose, Phys. Rev. B \textbf{66}, (2002) 161402


\bibitem{6}
  H. -S. Sim, H. -W. Lee, K. J. Chang, Phys. Rev. Lett. \textbf{87}, (2001) 096803

\bibitem{7}
  P. Havu, T. Torsti, M. J. Puska, R. M. Nieminen, Phys. Rev. B \textbf{66}, (2002) 075401


\bibitem{8}
  P. Khomyakov, G. Brocks, Phys. Rev. B \textbf{74}, (2006) 165416


\bibitem{ex6}
  R. Gutierrez, F. Grossmann, R. Schmidt, Acta Phys. Pol. B \textbf{32}, (2001) 443

\bibitem{ex2}
  Y. Egami, T. Ono, K. Hirose, Phys. Rev. B \textbf{72}, (2005) 125318


\bibitem{ex5}
  P. Major, V. G. Suarez, S. Sirichantaropass, J. Cserti, C. J. Lambert, J. Ferrer, G. Tichy, Phys. Rev. B
 \textbf{73} (2006) 045421

\bibitem{ex7}
  J. K. Viljas, J. C. Cuevas, F. Pauly, M. Hafner, Phys. Rev. B \textbf{72}, (2005) 245415

\bibitem{mirror}
  H. -W. Lee and C. S. Kim, Phys. Rev. B \textbf{63}, (2001) 075306


\bibitem{wannier1}
  N. Marzari, D. Vanderbilt, Phys. Rev. B \textbf{56}, (1997) 12847
 
\bibitem{wannier2}
 O. K. Andersen, T. Saha-Dasgupta, Phys. Rev. B \textbf{62}, (2000) R16219
 



\end{thebibliography}
\end{document}